\documentstyle[epsfig,12pt]{article}
\newcommand{\be}{\begin{eqnarray}}
\newcommand{\ee}{\end{eqnarray}}

\def\lsim{\mathrel{\rlap{\lower4pt\hbox{\hskip1pt$\sim$}}
    \raise1pt\hbox{$<$}}}               
\def\gsim{\mathrel{\rlap{\lower4pt\hbox{\hskip1pt$\sim$}}
    \raise1pt\hbox{$>$}}}               

\begin{document}

\begin{figure}[htb]

\epsfxsize=6cm \epsfig{file=logo_INFN.epsf}

\end{figure}

\vspace{-4.75cm}

\Large{\rightline{Sezione ROMA III}}
\large{
\rightline{Via della Vasca Navale 84}
\rightline{I-00146 Roma, Italy}
}

\vspace{0.6cm}

\rightline{INFN-RM3 98/4}
\rightline{September 1998}

\normalsize{}

\vspace{2cm}

\begin{center}

\Large{Light-Baryon Spectroscopy and the Electromagnetic Form Factors 
in the Quark Model\footnote{To appear in {\bf Few-Body Systems 
Supplementum}: Proceedings of the Joint ECT*/JLAB Workshop on {\em 
$N^*$ Physics and non-perturbative QCD}, ECT* (Italy), May 18-29, 1998, 
eds. V. Burkert, N. Mukhopadhyay, B. Saghai and S. Simula.}}

\vspace{0.5cm}

\large{F. Cardarelli$^1$, E. Pace$^2$, G. Salm\`e$^3$ and S. Simula$^1$}

\vspace{0.25cm}

\normalsize{$^1$INFN, Sezione Roma III, Via della Vasca Navale 84, 
I-00146 Roma, Italy\\ $^2$Dipartimento di Fisica, Universit\`a di Roma 
"Tor Vergata", and INFN, Sezione Tor Vergata, Via della Ricerca 
Scientifica 1, I-00133, Rome, Italy\\ $^3$INFN, Sezione di Roma I, P.le 
A. Moro 2, I-00185 Rome, Italy}

\end{center}

\vspace{0.5cm}

\begin{abstract}

\noindent The momentum distributions of the constituent quarks inside the
nucleon and the prominent electroproduced nucleon resonances are
investigated in the two most sophisticated, available quark potential
models, based respectively on the assumption of the valence + gluon
dominance and on the exchange of the pseudoscalar Goldstone-bosons arising
from the spontaneous breaking of chiral symmetry. It is shown that both
models predict a large, similar content of high-momentum components, due to
the short-range part of the interquark interaction, which affect the
behaviour of both elastic and transition electromagnetic form factors at
large values of the momentum transfer. The electromagnetic form factors are
calculated within a relativistic approach formulated on the light-front,
adopting a one-body current with constituent quark form factors. The results
suggest that soft, non-perturbative effects can play a relevant role for
explaining the existing data on elastic as well as transition form factors
(at least) for $Q^2 \lsim 10 \div 20 ~ (GeV/c)^2$.

\end{abstract}

\newpage

\pagestyle{plain}

\section{Introduction}

\indent The aim of this contribution is to address few relevant questions
concerning the possible consistency of the predictions of the constituent
quark ($CQ$) model with existing data on the electromagnetic (e.m.)
properties of the nucleon and the most prominent electroproduced nucleon
resonances at large values of the squared four-momentum transfer $Q^2$. In
these kinematical regions ($Q^2 \gsim$ few $(GeV/c)^2$) the $pQCD$ hard
scattering mechanism appears to be able to explain qualitatively existing
data (see, e.g., Ref. \cite{Stoler}), but results from $QCD$ sum rules (see,
e.g., Ref. \cite{Radyushkin}) seem to suggest that also the soft Feynman
mechanism can account for the same data as well.

\indent Since the high-$Q^2$ behaviour of the form factors is correlated
with the high-momentum tail of the $CQ$ momentum distribution in the nucleon
and its resonances, we first investigate the light-baryon wave functions
generated by the two most sophisticated, available quark potential models,
based respectively on the assumption of the valence + gluon dominance
\cite{Isgur} and on the exchange of the pseudoscalar Goldstone-bosons
arising from the spontaneous breaking of chiral symmetry \cite{Glozman}. It
will be shown that both models predict a large content of high-momentum
components due to the short-range part of the interquark interaction.
Moreover, despite the different behaviour of the two models at short
distances, the high-momentum tails of the light-baryon wave functions turn
out to be quite similar. These high-momentum components are known to affect
significantly the large-$Q^2$ behaviour of both elastic and transition e.m.
form factors, which are calculated adopting the light-front quark model of
Refs. \cite{nucleon,delta}. We will point out that: i) the introduction of
constituent quark form factors in the one-body e.m. current is essential in
order to explain the detailed $Q^2$ behaviour of the nucleon elastic data;
ii) the short-range spin-spin interaction generating the $N - \Delta(1232)$
mass splitting is also responsible for the faster-than-dipole fall-off of
the $N - \Delta(1232)$ magnetic transition form factor at large $Q^2$; iii)
an approximate dipole fall-off of the $N - S_{11}(1535)$ transition magnetic
form factor can be obtained, provided the nucleon elastic data are
reproduced. Our results suggest that soft, non-perturbative physics can
yield a relevant contribution for explaining the existing data on the
high-$Q^2$ behaviour of elastic and transition e.m. form factors, in accord
with the findings of $QCD$ sum rules.

\section{Quark potential models and the $CQ$ momentum distribution}

\indent The $CQ$ model is known to be a phenomenological model able to
explain the basic features of many static hadron properties, like the baryon
(and meson) mass spectra. Within this model the $CQ$'s are the only relevant
degrees of freedom in baryons, all the other degrees of freedom being frozen
in the $CQ$ mass ($m_i$) and interaction. The baryon wave function $\Psi_B$ 
is therefore eigenfunction of a Schroedinger-type equation, viz.
 \be
    \label{HB}
    \hat{H} \Psi_B = [\hat{T} + \hat{V}] \Psi_B =  M_B \Psi_B
 \ee
where $M_B$ is the baryon mass, $\hat{T} = \sum_{i=1}^3 \sqrt{|\vec{p}_i|^2
+ m_i^2}$ is the kinetic term and $\hat{V} = \hat{V}_{conf} +
\hat{V}_{s.r.}$ the interaction term, given by a long-range confining part
$\hat{V}_{conf}$ and a short-range component $\hat{V}_{s.r.}$ responsible
for the hyperfine mass splitting. The confining potential is usually derived
from a Lorentz-scalar interaction and, as suggested by the spectroscopy and
lattice $QCD$ calculations, it can be taken linearly dependent on the
quark-quark distance $r_{ij} \equiv |\vec{r}_i - \vec{r}_j|$, namely
$\hat{V}_{conf} \rightarrow \hat{V}_s = \sum_{i<j} b \cdot r_{ij}$, where
$b$ is the string tension. As for the short-range part of the interquark
potential, the most sophisticated choices existing in the literature are
based on two alternative mechanisms of boson exchange among $CQ$'s: the
one-gluon-exchange ($OGE$) model of Ref. \cite{Isgur} and the pseudoscalar
Goldstone-boson exchange ($GBE$) model of Ref. \cite{Glozman}.

\indent The semi-relativistic Hamiltonian model developed in Ref. \cite
{Isgur} is very successful in both meson and baryon sectors: it reproduces a
large amount of experimental masses and solves the so-called baryon
spin-orbit puzzle. The latter consists in the apparent absence of a
significant spin-orbit splitting in the light-baryon mass spectrum at
variance with naive expectations. The puzzle was solved by Isgur and
co-workers \cite{Isgur} by partially compensating the vector spin-orbit term
with the Thomas-Fermi precession spin-orbit term arising from the scalar
confining interaction, and by introducing (semi)relativistic corrections to
the interquark potential, which yield a significant suppression of the
interaction strength in case of light quarks. Nevertheless, a residual
problem still remains in the generally good picture given by the $OGE$
model: negative-parity states are below positive-parity ones, in clear
contrast to the observation.

\indent In the $GBE$ model of Ref. \cite{Glozman} the short-range part of
the $CQ$ interaction is generated by the exchange of the pseudoscalar
Goldstone bosons arising from the spontaneous breaking of chiral symmetry.
Such a potential model predicts baryon masses in quite good agreement with
the experimental data and, in particular, thanks to the flavour dependence
of the exchanged mesons, the $GBE$ model is able to yield the correct
ordering among positive and negative parity states. However, as pointed out
in Ref. \cite{ConfIII}, the agreement with the mass spectrum is obtained
only when the Thomas-Fermi precession spin-orbit term due to the scalar
confining interaction is (arbitrarily) neglected. Therefore the baryon
spin-orbit puzzle is still to be solved within the $GBE$ model (see for 
details \cite{ConfIII}). Despite the mentioned flaws, we stress that both
the $OGE$ and $GBE$ models yield a quite good overall description of the
light-baryon spectrum, although they remarkably differ at short interquark
distances.

\indent The wave equation (\ref{HB}) has been solved by expanding the wave
function $\Psi_B$ onto a (truncated) set of harmonic oscillator basis states
and applying to the Hamiltonian the Rayleigh-Ritz variational principle. We
have explicitly checked that a sufficiently large number of basis states has
been included in order to obtain the full convergence of the quantities
considered in this work. The $CQ$ momentum distribution $n(p)$, defined as
$n(p) = \int d\Omega_{\vec{p}} d\vec{p}_2 d\vec{p}_3 ~ \delta(\vec{p} +
\vec{p}_2 + \vec{p}_3) ~ |\Psi_B|^2$, calculated using the $OGE$ and $GBE$
models, is shown in Fig. 1 in case of the nucleon and compared with the
results obtained adopting only the (linear) confining part of the two
interactions. It can be seen that for both models the short-range part of
the potential produce a remarkable content of high-momentum components,
which turn out to be not very sensitive to the specific interaction model.

\begin{figure}[htb]

\parbox{7.75cm}{\epsfxsize=7.75cm \epsfig{file=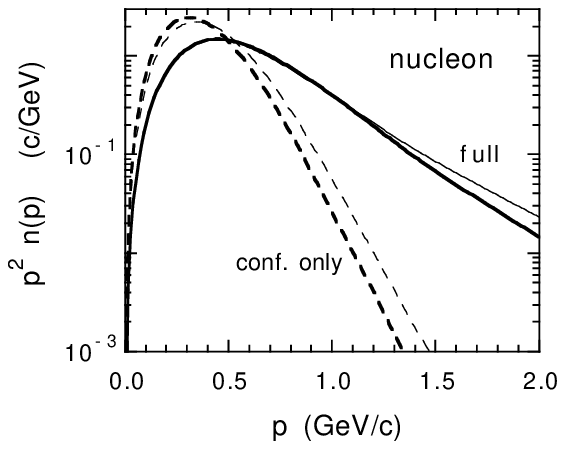}} \ $~~$ \
\parbox{7.75cm}{\small {\bf Figure 1}. $CQ$ momentum distribution $p^2 ~
n(p)$ in the nucleon versus the internal $CQ$ momentum $p$. Thick and thin
lines correspond to the $OGE$ and $GBE$ models of Refs. \cite{Isgur} and
\cite{Glozman}, respectively. The solid lines are the results obtained using
the full interaction models, whereas the dashed lines correspond to the case
in which only their (linear) confining parts are considered.}

\end{figure}

\indent The $CQ$ momentum distributions $n(p)$ in the $\Delta(1232)$,
$S_{11}(1535)$ and $F_{15}(1680)$ resonances, obtained within the $OGE$ and
$GBE$ models, are reported in Fig. 2 and compared with the ones in the
nucleon. It can clearly be seen that, although the $OGE$ and $GBE$ models
substantially differ at short interquark distances, the high-momentum tails
of the baryon wave functions are quite similar in both models with the only
(partial) exception of the $S_{11}(1535)$ resonance. Since the high-$Q^2$
behaviour of the form factors is qualitatively correlated with the
high-momentum tail of the $CQ$ momentum distribution, we naively expect the
same high-$Q^2$ behaviour for the elastic nucleon, $N - S_{11}(1535)$ and $N
- F_{15}(1680)$ transition form factors, while a faster fall-off is expected
in case of the $N - \Delta(1232)$ transition. Such features are indeed
present in the existing high-$Q^2$ data, namely both the elastic nucleon,
the $N - S_{11}(1535)$ and the $N - F_{15}(1680)$ transition magnetic form
factors exhibit approximately the same dipole fall-off for $Q^2$ greater
than few $(GeV/c)^2$, while the $N - \Delta(1232)$ transition magnetic form
factor drops faster than a dipole (see \cite{Stoler}).

\begin{figure}[htb]

\parbox{7.75cm}{\epsfxsize=7.75cm \epsfig{file=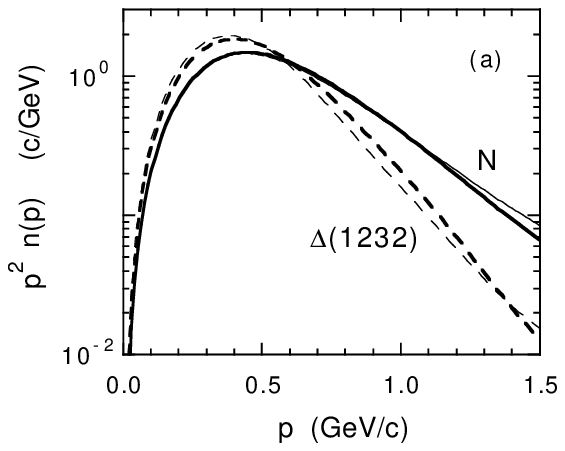}} \ $~~$ \
\parbox{7.75cm}{\small {\bf Figure 2}. $CQ$ momentum distributions $p^2 ~
n(p)$ in the $\Delta(1232)$ (a), $S_{11}(1535)$ (b) and $F_{15}(1680)$ (c)
resonances (dashed lines). Thick and thin lines correspond to the $OGE$
\cite{Isgur} and $GBE$ \cite{Glozman} potential models, respectively. For
comparison in each picture the $CQ$ momentum distributions in the nucleon
(solid lines) are explicitly shown.}

\vspace{0.25cm}
 
\parbox{7.75cm}{\epsfxsize=7.75cm \epsfig{file=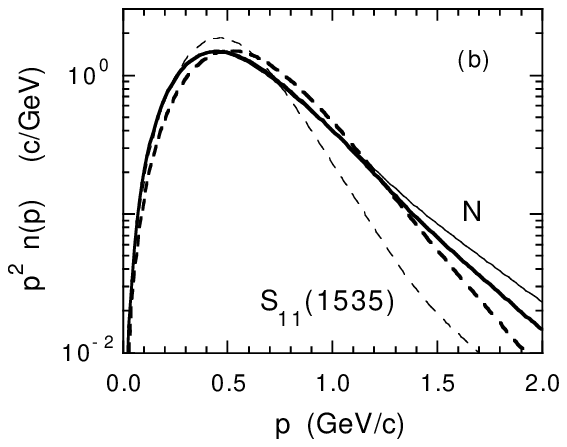}} \ $~~$ \
\parbox{7.75cm}{\epsfxsize=7.75cm \epsfig{file=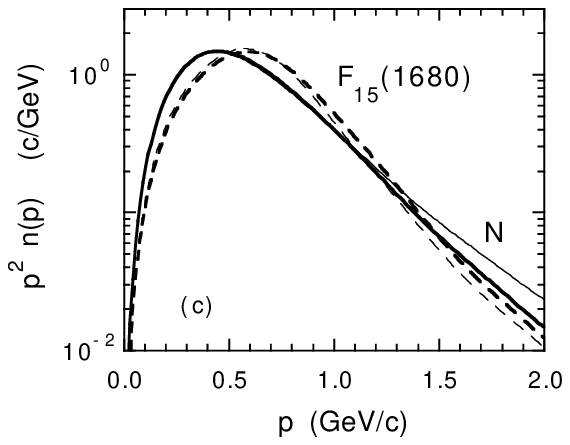}}

\end{figure}

\indent In the next Section the issue of the $CQ$ model predictions for the 
nucleon elastic and transition e.m. form factors will be addressed and to 
this end the relativistic quark model of Ref. \cite{nucleon}, formulated on 
the light front, is adopted. Since the high-momentum components generated by 
the $OGE$ and $GBE$ models turn out to be quite similar, in what follows we 
will limit ourselves to consider explicitly the light-baryon wave functions 
of the $OGE$ model only.

\section{Elastic and transition e.m. form factors}

\indent The effects of the high-momentum tail of the nucleon wave function
generated by the $OGE$ model on the elastic nucleon e.m. form factors have
been investigated for the first time in Ref. \cite{nucleon}, where a
relativistic one-body e.m. current was adopted. It was shown that both the
relativistic effects and the high-momentum components of the wave function
lead to a sizeable overestimation of the proton form factors both at low
($\lsim 1 ~ (GeV/c)^2$) and high ($\gsim 1 ~ (GeV/c)^2$) $Q^2$. This result
is not surprising, because a pure valence quark model (i.e., without any
effect from sea quark pairs) is not expected to describe dynamical
properties like the e.m. form factors. One could argue that, in order to
keep safe the $CQ$ picture of the hadron structure, the $CQ$ itself can be
viewed as a non-elementary object whose structure takes into account in an
effective way the presence of non-valence components. Thus, in Ref.
\cite{nucleon} a one-body e.m. current with $CQ$ form factors was adopted.
The latter ones cannot be derived directly from $QCD$ and therefore one is
limited to constrain the $CQ$ form factors by the request of reproducing the
nucleon elastic data and to ask if existing data on the transition form
factors are consistent with the same one-body e.m. current. This program has
been partially carried out in Refs. \cite{nucleon,delta}: adopting the
baryon wave functions of the $OGE$ model, the $CQ$ form factors were firstly
fixed through the reproduction of the nucleon elastic data and then used to
calculate {\em without free parameters} the $N - \Delta(1232)$ transition
form factors, obtaining a good overall description of the existing
data\footnote{Due to the limitations imposed by the violation of the
so-called angular condition (see Ref. \cite{delta}), we will consider in this 
work only the predictions of our light-front model for the dominant magnetic 
transition form factor.}.

\indent In this contribution we want to point out that, once the elastic
data are reproduced, the faster-than-dipole fall-off of the $N -
\Delta(1232)$ magnetic transition form factor can be obtained only when the
effects from the short-range spin-spin interaction are taken into account in
the baryon wave functions. To this end we have calculated the nucleon
elastic form factors adopting two different wave functions obtained from the
full $OGE$ interaction and from its (linear) confining part only (see solid
and dashed lines in Fig. 1, respectively). For each wave functions the $CQ$
form factors have been determined by the request of reproducing the nucleon
(and pion) data. The results are reported in Fig. 3 and it can be clearly
seen that: ~ i) the introduction of the $CQ$ form factors in the one-body
e.m. current is essential in order to explain the detailed $Q^2$ behaviour
of the nucleon elastic data, and ~ ii) once appropriate $CQ$ form factors
are introduced, the nucleon data alone cannot distinguish between models
with and without short-range interaction effects. It should be mentioned
that the phenomenological $CQ$ form factors associated to the full $OGE$
wave function and to the much softer wave function generated by the linear
confining interaction, correspond to quite different values of the $CQ$
size, namely $\sim 0.5$ and $\sim 0.2 ~ fm$, respectively.

\begin{figure}[htb]

\vspace{0.25cm}

\parbox{7.75cm}{\epsfxsize=7.75cm \epsfig{file=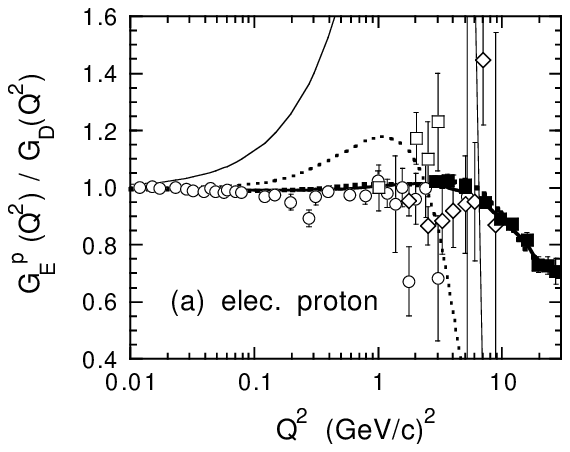}} \ $~~$ \
\parbox{7.75cm}{\epsfxsize=7.75cm \epsfig{file=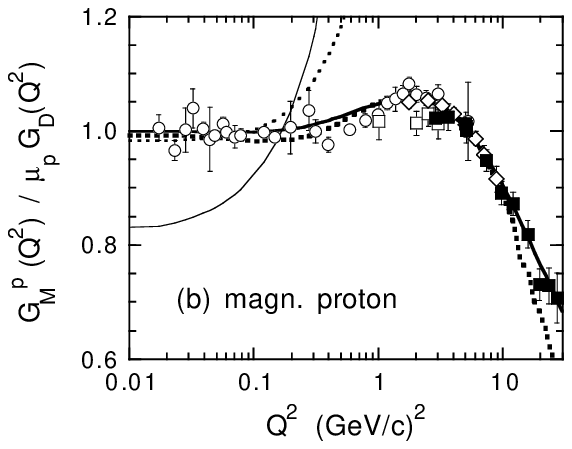}}

\vspace{0.25cm}

\parbox{7.75cm}{\epsfxsize=7.75cm \epsfig{file=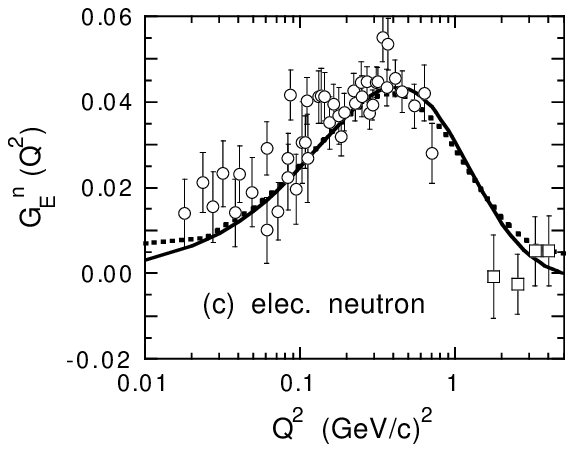}} \ $~~$ \
\parbox{7.75cm}{\epsfxsize=7.75cm \epsfig{file=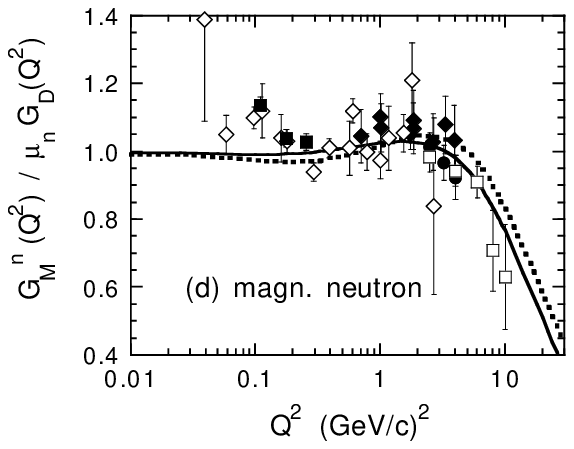}}

\vspace{0.25cm}

\parbox{0.25cm}{~~} \ $~~$ \parbox{15.5cm}{\small {\bf Figure 3}. Nucleon 
form factors $G_E^p(Q^2) / G_D(Q^2)$ (a), $G_M^p(Q^2) / \mu_p G_D(Q^2)$ (b),
$G_E^n(Q^2)$ (c) and $G_M^n(Q^2) / \mu_n G_D(Q^2)$ (d) versus $Q^2$. The
solid and dotted thick lines correspond to the results obtained using the
nucleon wave functions resulting from the full $OGE$ model \cite{Isgur} and
from its (linear) confining part only, including $CQ$ form factors in the
one-body e.m. current. In (a) and (b) the dotted and solid thin lines
correspond to the case in which point-like $CQ$'s are assumed. Data are
quoted in details in Ref. \cite{nucleon}. The dipole form is given by
$G_D(Q^2) = 1 / (1 + Q^2 / 0.71)^2$.}

\end{figure}

\indent In Fig. 4 our parameter-free predictions for the $N - \Delta(1232)$ 
magnetic transition form factor are compared with existing data. It can 
clearly be seen that the effects from the short-range part of the $CQ$ 
interaction, which is responsible for the $N - \Delta(1232)$ mass splitting 
and also for the different high-momentum tails of the $N$ and $\Delta(1232)$ 
wave functions (see Fig. 2(a)), are now essential in order to reproduce the 
faster-than-dipole fall-off of the transition form factor at large $Q^2$ 
\footnote{The deviations from the data at low $Q^2$ are discussed in
Ref. \cite{delta}. For the photon point see also Ref. \cite{GDH}.}.

\begin{figure}[htb]

\vspace{0.25cm}

\parbox{7.75cm}{\epsfxsize=7.75cm \epsfig{file=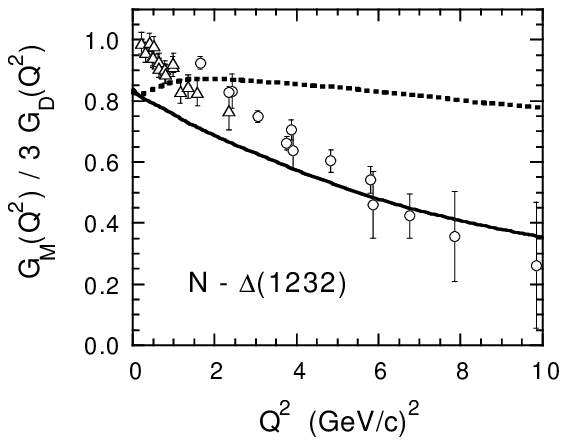}} \ $~~$ \
\parbox{7.75cm}{\small {\bf Figure 4}. The $N - \Delta(1232)$ transition
magnetic form factor $G_M(Q^2) / 3G_D(Q^2)$ versus $Q^2$. The solid and
dotted lines correspond to the results obtained using the $N$ and
$\Delta(1232)$ wave functions resulting from the full $OGE$ model
\cite{Isgur} and from its (linear) confining part only. The $CQ$ form 
factors are the ones used to reproduce the nucleon form factors (see
Fig. 3). Triangles and open dots are from Ref. \cite{ND}(a) and (b),
respectively.}

\end{figure}

\indent Finally, in Fig. 5 the parameter-free predictions for the $N - 
S_{11}(1535)$ magnetic transition form factor $G_M^*(Q^2) / G_D(Q^2)$, 
obtained in Ref. \cite{S11} using the full $OGE$ wave functions, are 
reported and compared with available inclusive electroproduction data 
\cite{Stoler}. It can be seen that the naive expectation of a dipole 
fall-off at large $Q^2$, based on the similar high-momentum behaviours 
of the $N$ and $S_{11}(1535)$ wave functions (see Fig. 2(b)), is fully 
confirmed by the explicit calculations. Therefore, we point out that 
the predictions of our light-front $CQ$ model for the transition form 
factors to the most prominent electroproduced nucleon resonances are 
not inconsistent with the data at large $Q^2$, suggesting that soft, 
non-perturbative effects can still play a decisive role in the 
nucleon-resonance transition form factors at least up to $Q^2 \sim 10 
\div 20 ~ (GeV/c)^2$, in accord with the results of $QCD$ sum rules 
\cite{Radyushkin}. 

\begin{figure}[htb]

\parbox{7.75cm}{\epsfxsize=7.75cm \epsfig{file=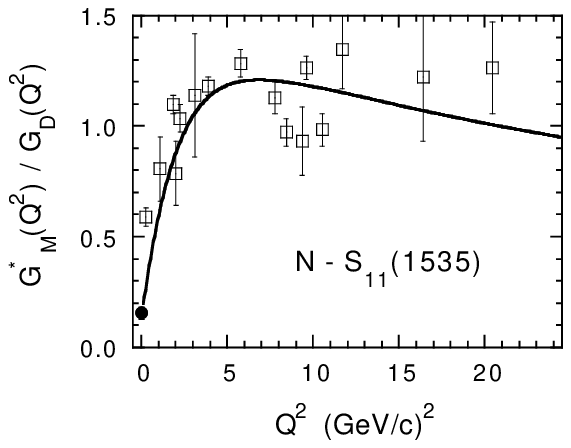}} \ $~~$ \
\parbox{7.75cm}{\small {\bf Figure 5}. The $N - S_{11}(1535)$ transition
magnetic form factor $G_M^*(Q^2) / G_D(Q^2)$ versus $Q^2$. The solid line
corresponds to the results obtained using the $N$ and $S_{11}(1535)$ wave
functions resulting from the full $OGE$ model \cite{Isgur}. The $CQ$ form 
factors are the ones used to reproduce the nucleon form factors (see
Fig. 3). Electroproduction data are from Ref. \cite{Stoler}, while the 
photon point is from Ref. \cite{PDG96}.} 

\end{figure}

\section{Conclusions}

\indent The momentum distributions of the constituent quarks inside the
nucleon and the prominent electroproduced nucleon resonances have been
investigated in the two most sophisticated, available quark potential
models, based respectively on the assumption of the valence + gluon
dominance and on the exchange of the pseudoscalar Goldstone-bosons arising
from the spontaneous breaking of chiral symmetry. It has been shown that
both models predict a large, similar content of high-momentum components due
to the short-range part of the interquark interaction. Elastic and
transition e.m. form factors have been calculated within a relativistic
approach formulated on the light-front, adopting a one-body current with
constituent quark form factors. The main results are: i) the introduction of
constituent quark form factors is essential in order to explain the detailed
$Q^2$ behaviour of the nucleon elastic data; ii) the short-range spin-spin
interaction generating the $N - \Delta(1232)$ mass splitting is also
responsible for the faster-than-dipole fall off of the $N - \Delta(1232)$
magnetic transition form factor at large $Q^2$; iii) an approximate dipole
fall-off of the $N - S_{11}(1535)$ transition magnetic form factor can be
obtained, provided the nucleon elastic data are reproduced. Our results
suggest that soft, non-perturbative physics can yield a relevant, decisive
contribution for explaining the existing data on the nucleon elastic as well
as transition e.m. form factors (at least) up to $Q^2 \sim 10 \div 20 ~
(GeV/c)^2$.

\end{document}